\documentclass[aps,prl,reprint,amsmath,amssymb,citeautoscript,superscriptaddress]{revtex4-1}

\usepackage{amsmath} 
\usepackage{graphicx} 
\usepackage[T1]{fontenc}

\newcommand{\supp}{See Supplemental Material at [url], which includes Refs.~[32]--[44], for additional details on sample
characterization, optical microscopy, spectroscopy, and theoretical and computational methodology}

\begin{document}

\title{Exciton Binding Energy and Nonhydrogenic Rydberg Series in Monolayer WS$_2$}

\author{Alexey Chernikov}
\email{aac2183@columbia.edu}
\affiliation{Departments of Physics and Electrical Engineering, Columbia University, 538 West 120th Street, New York, NY 10027, USA}

\author{Timothy C. Berkelbach}
\affiliation{Department of Chemistry, Columbia University, 3000 Broadway, New York, NY 10027, USA}

\author{Heather M. Hill} \author{Albert Rigosi} \author{\mbox{Yilei Li}} \author{\"Ozgur B. Aslan} 
\affiliation{Departments of Physics and Electrical Engineering, Columbia University, 538 West 120th Street, New York, NY 10027, USA}

\author{David R. Reichman} 
\affiliation{Department of Chemistry, Columbia University, 3000 Broadway, New York, NY 10027, USA}

\author{Mark S. Hybertsen} 
\affiliation{Center for Functional Nanomaterials, Brookhaven National Laboratory, Upton, NY 11973-5000, USA}

\author{Tony F. Heinz}
\email{tony.heinz@columbia.edu}
\affiliation{Departments of Physics and Electrical Engineering, Columbia University, 538 West 120th Street, New York, NY 10027, USA}

\begin{abstract} 
We have experimentally determined the energies of the ground and first four excited excitonic states of the fundamental optical
transition in monolayer WS$_2$, a model system for the growing class of atomically thin two-dimensional semiconductor crystals.
From the spectra, we establish a large exciton binding energy of 0.32 eV and a pronounced deviation from the usual hydrogenic
Rydberg series of energy levels of the excitonic states. We explain both of these results using a microscopic theory in which the
non-local nature of the effective dielectric screening modifies the functional form of the Coulomb interaction. These strong but
unconventional electron-hole interactions are expected to be ubiquitous in atomically thin materials.
\end{abstract}

\maketitle

Atomically thin materials such as graphene and transition metal dichalcogenides~(TMDs) exhibit remarkable physical properties
resulting from their reduced dimensionality~\cite{Novoselov2005}.  The family of TMDs is an especially promising platform for
fundamental studies of two-dimensional (2D) systems, with potential applications in optoelectronics and valleytronics due to their
direct gap, semiconducting nature in the monolayer limit~\cite{Mak2010,Li2007,Splendiani2010,Britnell2013,Ross2013,Zhao2013}.  The
recent advances in this emerging field include strongly enhanced photoluminescence~\cite{Mak2010,Splendiani2010}, efficient
spin-valley coupling~\cite{Mak2012a,Zeng2012,Cao2012,Jones2013}, pronounced many-body effects~\cite{Mak2012,Ross2013}, and
high-performance in field-effect transistors~\cite{Lembke2012}.

The 2D character of monolayer TMDs suggests a strong enhancement of the Coulomb interaction.  The resulting formation of bound
electron-hole pairs, or excitons, can dominate the optical and charge-transport properties~\cite{Haug2009}.  A microscopic
understanding of how excitons are formed from otherwise free carriers is critical both for the elucidation of the underlying
many-body physics in such materials and for their use in electronic and photonic devices, since the response of charged free
carriers and neutral excitons to applied fields differs dramatically.  While theoretical and computational studies have predicted
exciton binding energies as high as 1~eV~\cite{Ramasubramaniam2012,Komsa2012,Shi2013,Qiu2013,Berghauser2014}, a direct measurement
of the exciton binding energy in atomically thin TMDs is still lacking (however, see Note added).

In this work we experimentally and theoretically investigate the properties of excitons in mono- and few-layer TMDs, identifying
and characterizing not only the ground-state exciton, but the full sequence of excited (Rydberg) exciton states.  Analyzing our
sensitive measurements of the optical reflection spectra of these materials, both empirically and using a physically motivated
model for the non-local screening in TMDs, results in an estimate of 0.32($\pm$0.04) eV for the 1s exciton binding energy and
2.41($\pm$0.04) eV for the quasiparticle gap of monolayer WS$_2$.  Remarkably, we also find significant deviations from the
conventional hydrogenic model typically employed for the description of Wannier excitons in inorganic
semiconductors~\cite{Klingshirn2007}, and explain our findings in terms of a microscopic theory that highlights the peculiar form
taken by the electron-hole interaction in this class of novel materials~\cite{Keldysh1979,Cudazzo2011,Berkelbach2013}. 

The specific material studied here is WS$_2$, a representative member of the TMD family that includes MoS$_2$, MoSe$_2$, and
WSe$_2$, all of which share similar properties with respect to atomic and electronic structure.  The advantage of WS$_2$ for this
study is the large spin-orbit splitting between the A and B excitons of about 0.4~eV~\cite{Zhao2013}, allowing for a study of the
low-energy excitons unobscured by features from higher-lying transitions.  In addition, the electronic transitions in the WS$_2$
samples exhibit narrow spectral features, permitting identification and analysis of many excited excitonic states and detailed
quantitative comparison with theoretical predictions. Sample preparation and characterization details can be found in the
Supplemental Material~\footnote{\supp}.

\begin{figure}[t] 
\centering 
\includegraphics[width=8.3 cm]{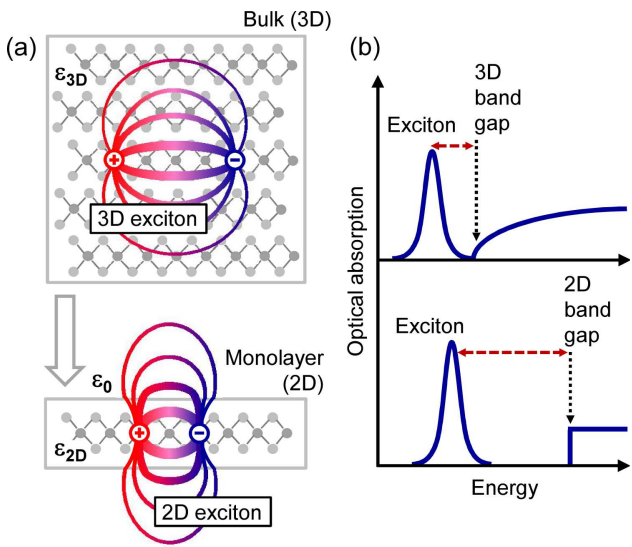}
\caption{
(a) Real-space representation of electrons and holes bound into excitons for the three-dimensional bulk and a
quasi-two-dimensional monolayer.  The changes in the dielectric environment are indicated schematically by different dielectric
constants $\varepsilon_{3D}$, $\varepsilon_{2D}$, and the vacuum permittivity $\varepsilon_{0}$.  (b) Impact of the dimensionality
on the electronic and excitonic properties, schematically represented by optical absorption.  The transition from 3D to 2D is
expected to lead to an increase of both the band gap and the exciton binding energy (indicated by the red dashed line).  The
excited excitonic states and Coulomb correction for the continuum absorption have been omitted for clarity.}
\label{fig1} 
\end{figure}

Experimental and theoretical studies to date have clearly demonstrated that the basic excitonic properties of a three-dimensional
bulk semiconductor differ fundamentally from those of a 2D monolayer of the same material.  The real-space origin of this behavior
in TMDs is illustrated schematically in Fig.~\ref{fig1}(a).  In contrast to bulk, the  electron and hole forming an exciton in
monolayer TMDs are strongly confined to the plane of the monolayer and additionally experience reduced screening due to the change
in the dielectric environment.  These effects have two major implications for the electronic and excitonic properties of the
material, shown by a schematic representation of the optical absorption in Fig.~\ref{fig1}(b).  First, the quasiparticle band gap
is expected to increase for the monolayer. Second, the enhanced electron-hole interaction is expected to increase the exciton
binding energy.  In the absence of dielectric effects this yields an exciton binding energy that is a factor of four larger in 2D
than in 3D.  In the limit of atomically thin materials, however, the dielectric screening is also reduced because the electric
field lines joining the electron and hole begin to extend outside of the sample as shown in Fig.~\ref{fig1}(a), potentially
yielding an even greater enhancement factor.  This so-called ``dielectric confinement'' or ``image charge
effect''~\cite{Keldysh1979,Hanamura1988} was observed in nano-structured materials such as single-walled carbon
nanotubes~\cite{Deslippe2009} and layered organic-inorganic perovskites~\cite{Tanaka2005}.  The effectiveness of the dielectric
screening thus depends on the separation between the electron and hole, giving rise to a non-local dielectric screening. This
modifies the form of the interaction potential~\cite{Keldysh1979,Hanamura1988,Cudazzo2011,Berkelbach2013} and causes a significant
change of the disposition of the energies of the excitonic states, as discussed in more detail below.

To access these exciton properties experimentally we study the so-called excitonic Rydberg series, i.e.,~the excited states of the
bound electron-hole pairs, labeled in analogy to the hydrogen series as 2s, 3s, and so on.  In contrast to p- or d-like states
with nonzero orbital angular momentum, these transitions are dipole-allowed and are thus found in the linear optical spectra of
many semiconductors with peak positions located between the quasiparticle band gap and the exciton 1s ground
state~\cite{Klingshirn2007,Haug2009}.  The energy separation of these resonances corresponds to a hydrogenic progression for
Wannier-like excitons.  In addition, the coupling of the excited states to light is reduced compared to the main transition so
that their spectral weight decreases with increasing quantum number.

\begin{figure}[b] 
\centering 
\includegraphics[width=7 cm]{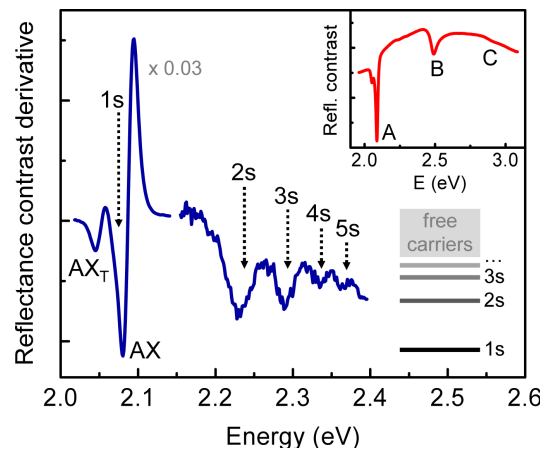} 
\caption{
The derivative of the reflectance contrast spectrum $\frac{d}{dE}(\Delta R/R)$ of the WS$_{2}$ monolayer.  The exciton ground
state and the higher excited states are labeled by their respective quantum numbers (schematically shown at bottom-right).  The
spectral region around the 1s transition (AX) and the trion peak (AX$_T$) of the A exciton is scaled by a factor of 0.03 for
clarity.  Inset shows the as-measured reflectance contrast $\Delta R/R$ for comparison, allowing for the identification of the A,
B, and C transitions.
} 
\label{fig2} 
\end{figure}

\begin{figure}[t] 
\centering 
\includegraphics[width=8.5 cm]{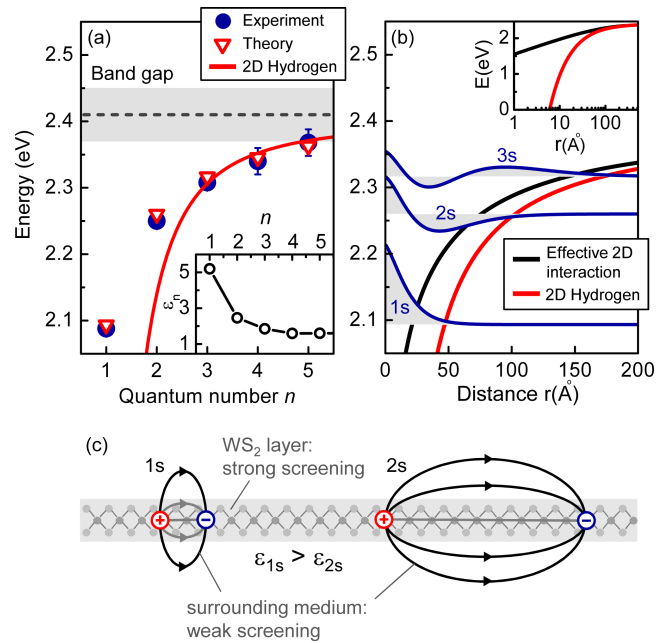} 
\caption{
(a)~Experimentally and theoretically obtained transition energies for the exciton states as a function of the quantum number $n$.
The fit of the $n=3,4,5$ data to the 2D hydrogen model for Wannier excitons is shown for comparison.  Grey bands represent
uncertainty in the quasi-particle band gap from the fitting procedure.  Corresponding effective dielectric constants are shown in
the inset.  (b)~Screened 2D interaction Eq.~(\ref{eq:interaction}) used in the model Hamiltonian (black) compared to the 2D
hydrogen interaction $1/r$ (red); a semilogarithmic plot is given in the inset.  Also shown are the corresponding energy levels
and radial wavefunctions up to $n=3$.  (c)~Schematic representation of electron-hole pairs forming 1s and 2s excitonic states in a
non-uniform dielectric environment.
} 
\label{fig3} 
\end{figure}

In our experiments we measure the reflectance contrast $\Delta R/R~=~(R_{\rm sample}-R_{\rm substrate})/R_{\rm substrate}$ of the
WS$_2$ monolayer sample at a temperature of 5~K. The experimental details are given in the Supplemental
Material~\footnotemark[\value{footnote}].  The spectrum, plotted in the inset of Fig.~\ref{fig2}, exhibits several pronounced
peaks on a broad background, the latter arising from interference effects induced by the 300-nm thick SiO$_2$ layer between the
sample and the Si substrate~\cite{Ross2013}.  The main transitions correspond to the so-called A, B, and C excitons in
WS$_2$~\cite{Zhao2013}.  A small additional feature on the low-energy side of the A peak is identified as a charged exciton (or
trion), with a binding energy on the order of 20--30 meV.  Such a feature has been observed in monolayers of other TMDs at low
temperatures~\cite{Mak2012,Ross2013} and indicates the presence of some unintentional residual doping in the WS$_2$ sample.  Here,
we focus on the properties of the A exciton, related to the fundamental band gap of the material.  In order to highlight the
otherwise weak signatures of the higher-lying excitonic transitions, we plot in Fig.~\ref{fig2} the derivative of the reflectance
contrast $\frac{d}{dE}(\Delta R/R)$ in the energy range of interest.  On the high-energy side of the exciton 1s ground state, we
observe multiple additional peaks, which we identify as the 2s, 3s, 4s, and 5s states of the A exciton, since the decrease of both
the peak intensity and the energy spacing for increasing energy are characteristic features of an excitonic Rydberg
series~\cite{Klingshirn2007,Haug2009}.  The peak positions extracted by taking the respective points of inflection, corresponding
to the zero-crossings of the second derivative, are plotted in Fig.~\ref{fig3}(a).  The respective energies are further confirmed
by simulating the material response with a multiple-Lorentzian fit (see Supplemental Material~\footnotemark[\value{footnote}]).

To calculate the exciton binding energy, we must first determine the quasiparticle band gap corresponding to the energy of a
separated electron-hole pair.  This is typically accomplished in semiconductors by fitting the excitonic peaks to a hydrogenic
Rydberg series~\cite{Klingshirn2007}.  In 2D, this hydrogen model employs an effective mass Hamiltonian, $H = -\hbar^2
\nabla_{\mathbf{r}}^2/2\mu + V_{eh}(r)$, where $\mu = 1/(m_e^{-1}+m_h^{-1})$ is the exciton reduced mass and $V_{eh}(r) =
-e^2/\varepsilon r$ is a locally-screened attractive electron-hole interaction.  This model predicts exciton transition energies
of $E_g - E_b^{(n)}$, where $E_g$ is the quasiparticle gap and \begin{equation} E_b^{(n)} = \frac{\mu
e^4}{2\hbar^2\varepsilon^2\left(n-1/2\right)^2} \end{equation} is the binding energy of the $n$th excitonic state.  In contrast,
the exciton energies seen in our experiments exhibit a much weaker scaling with the quantum number $n$, precluding a simple fit to
the data based on this model.  However, we observe that the $n=3-5$ peaks are reasonably hydrogenic and by fitting to these data
points only, we extract a quasiparticle band gap of $E_g = 2.41(\pm0.04)$~eV, where the error bars originate from the fitting
procedure.  Subtracting the 1s transition energy of 2.09~eV from this band gap, we find an exciton binding energy of $E_b =
0.32(\pm0.04)$~eV.

To provide insight into the non-hydrogenic physics of the $n=1,2$ excitons and the justification of a hydrogenic fit to the
$n=3-5$ excitons, we first consider the use of an effective dielectric constant in the hydrogenic Hamiltonian.  Using an exciton
reduced mass of $\mu = 0.16~m_0$ (as determined by density functional theory at the $K$ or $K^\prime$
point~\cite{Xiao2012,Berkelbach2013}, see Supplemental Material~\footnotemark[\value{footnote}]) and the quasiparticle band gap of
$E_g = 2.41$~eV, we determine the $n$-dependent dielectric constant $\varepsilon_n$ required to reproduce the experimental binding
energy of the $n$th exciton $E_{b,\textrm{exp}}^{(n)}$, i.e., $\varepsilon_n = [2\hbar^2 E_{b,\textrm{exp}}^{(n)}(n-1/2)^2/\mu
e^4]^{-1/2}$.  The results plotted in the inset of Fig.~\ref{fig3}(a) show a strong decrease in this effective dielectric constant
with increasing quantum number $n$.  Because the exciton radius increases with $n$, we conclude that the physically correct
electron-hole interaction is more strongly screened at short range, but only weakly screened at long range.  In particular, the
effective dielectric is nearly constant for $n=3-5$ (justifying our empirical use of the 2D hydrogen model for these data points),
but shows significant deviations for $n=1,2$.  This can be understood qualitatively in terms of a non-uniform dielectric
environment schematically illustrated in Fig.~\ref{fig3}(c).  The electric field between an electron and a hole forming an exciton
permeates both the thin layer of material with comparably strong screening and the surrounding medium with much weaker screening.
As the spatial separation between the charges increases, a larger portion of the electric field is located in the surrounding
low-dielectric medium and the effective screening is reduced.  This phenomenon of ``anti-screening,'' giving rise to
non-hydrogenic exciton behavior, has previously been predicted in carbon nanotubes, a quasi one-dimensional
semiconductor~\cite{Deslippe2009}.

To understand this behavior quantitatively, we apply our recently developed theory of excitons in transition metal
dichalcogenides~\cite{Berkelbach2013}.  The treatment is again based on a 2D effective mass Hamiltonian, but with a
nonlocally-screened electron-hole interaction described by the potential 
\begin{equation}\label{eq:interaction} 
V_{eh}(r) = -\frac{\pi e^2}{2r_0} \left[ H_0\left(\frac{r}{r_0}\right) - Y_0\left(\frac{r}{r_0}\right) \right], 
\end{equation} 
where $H_0$ and $Y_0$ are Struve and Bessel functions.  This interaction form describes the electrostatic interaction of two
charges within a thin 2D dielectric continuum~\cite{Keldysh1979,Hanamura1988,Cudazzo2011}.  The screening length $r_0$, which can
be related to the 2D polarizability of the monolayer material~\cite{Cudazzo2011}, gives a crossover length scale between a $1/r$
Coulomb interaction at large separation and a weaker $\log(r)$ interaction at small separation.  This modified functional form of
the interaction, which is a manifestation of the strong dielectric contrast between the monolayer WS$_2$ and its surroundings, is
responsible for the altered disposition of the low-lying excitonic states observed experimentally.

Using the above functional form for the screened interaction we have numerically calculated the radially symmetric, s-type
eigenstates of the excitonic Hamiltonian, again using the calculated exciton reduced mass $\mu = 0.16\ m_0$.  Taking only the
screening length $r_0$ and the band gap $E_g$ as free parameters, we find that we can very accurately fit the entire $n=1-5$
series of exciton levels with the values $r_0 = 75$ \AA\ and $E_g = 2.41$ eV.  These are the parameters which minimize the
root-mean-squared deviation between theory and experiment.  For this choice of parameters, the 1s exciton binding energy is found
to be 0.32~eV, and so both the band gap and the binding energy are found to agree with the values determined above.  We emphasize
that the adopted screening length should be understood as one that partially accounts for additional screening due to the
substrate, such that the intrinsic binding energy of WS$_2$ is expected to be larger than the value found here, in qualitative
agreement with \textit{ab initio} calculations~\cite{Ramasubramaniam2012,Komsa2012,Shi2013} (see Supplemental Material for a
discussion of the microscopic origin of the precise value of $r_0$~\footnotemark[\value{footnote}]).  Fig.~\ref{fig3}(b) depicts
the noticeably weakened interaction at small electron-hole separations, along with the first three calculated radial
wavefunctions.  The exciton radius is calculated to be approximately 30~\AA\ for the 1s exciton and even larger for the
higher-lying excitons, which supports a strictly 2D treatment when compared to the monolayer width of about 6 \AA.  Similarly,
this relatively large in-plane spatial extent implies a narrow reciprocal space distribution, justifying an effective mass
approximation centered around the $K$ and $K^\prime$ points of the Brillouin zone.  The above success of fitting to a hydrogenic
model is also explained by the present microscopic approach because the $n=3-5$ exciton wavefunctions are large enough in spatial
extent to predominantly probe the asymptotic $1/r$ form of the potential given in Eq.~(\ref{eq:interaction}).

\begin{figure}[t] 
\centering 
\includegraphics[width=8.5 cm]{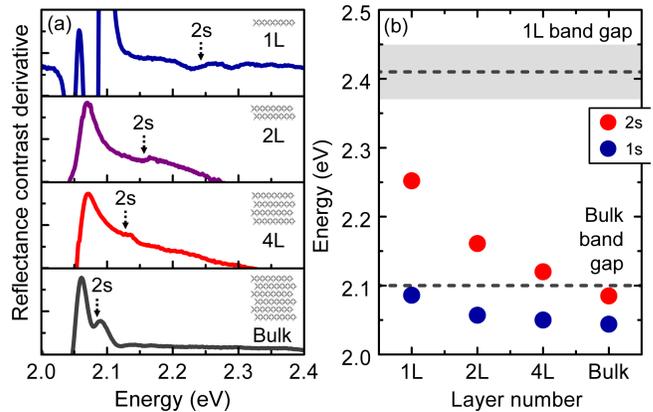} 
\caption{
(a) The derivative of the reflectance contrast spectra for the WS$_{2}$ monolayer~(1L), bilayer~(2L), tetralayer~(4L) and bulk.
The positions of the 2s exciton resonance are indicated by dotted arrows.  (b)~Energies of the 1s and 2s states for various layer
thicknesses.  Band gaps of the bulk and the monolayer are represented by the dashed lines.} 
\label{fig4} 
\end{figure}

Finally, to study the influence of the material thickness we monitor the spectral position of the 2s resonance for varying
thickness of the WS$_2$ sample.  Individual derivatives of the reflection contrast are plotted in Fig.~\ref{fig4}(a) for the
monolayer~(1L), bilayer~(2L), tetralayer~(4L), and bulk.  The corresponding energies of the 1s and 2s transitions are shown in
Fig.~\ref{fig4}(b), with higher exited states masked by additional spectral broadening.  Unlike for the case of the monolayer, the
bulk excitons are accurately treated with an anisotropic 3D hydrogenic Hamiltonian~\cite{Baldereschi1970} that accounts for
anisotropy in both the electron and hole band masses and in the dielectric tensor.  Using \textit{ab initio} calculated values, we
obtain a bulk exciton binding energy of 0.05~eV (see Supplemental Material~\footnotemark[\value{footnote}]), implying a band gap
of $E_g~=~$2.10~eV, both of which are in agreement with literature results for bulk WS$_2$~\cite{Beal1976}.  As the layer
thickness decreases, the 2s resonance shifts to higher energies, while the 1s resonance remains relatively unchanged, implying a
strong increase in both the exciton binding energy and the quasiparticle band gap.  Both shifts are found to be large in absolute
energies, but opposite in sign.  This explains the small change in the transition energy of the exciton ground state, similar to
findings reported for quasi-one-dimensional systems~\cite{Deslippe2009}.

The large binding energy of 0.32~eV and the non-hydrogenic behavior of the intra-excitonic states in monolayer WS$_2$ are expected
to be features common to other TMD materials, based on the strong similarity in their electronic structure.  Even larger values of
the binding energy may be expected for suspended and undoped layers, although the studied system represents the typical scenario
for experimental investigations.  These energies lead to a high thermal stability of excitons and combined with their high
oscillator strength should facilitate the application of these materials in photonic and excitonic devices.  The non-hydrogenic
series of the excited states resulting in large energy spacings between the individual states is expected to allow for efficient
exploitation of the intra-excitonic processes with potential applications in the far-IR and THz spectral range.  High-density
effects scaling with the strength of the Coulomb scattering such as screening and band-gap renormalization, generally important
for devices operating in this regime, e.g., lasers and concentrator solar-cells, can be potentially harvested for optical
modulators, saturable absorbers, and tunable emitters operating in the limit of atomic thickness.  Finally, the observed
properties of the WS$_2$ excitons render the material system highly suitable as a link between inorganic semiconductors with
spatially extended, weakly-bound Wannier excitons and organic semiconductors with spatially localized, strongly-bound Frenkel
excitons.  The presence of the strong Coulomb interaction opens up possibilities for both fundamental studies of the many-body
physics in 2D materials as well as for distinctive applications in optoelectronic devices. 

This work was supported in part by the Center for Re-Defining Photovoltaic Efficiency through Molecule Scale Control, an Energy
Frontier Research Center funded by the U.S. Department of Energy, Office of Science, Office of Basic Energy Sciences under Award
No.~DE-SC0001085.  Support for the reflection contrast measurements measurements was also provided by the National Science
Foundation through grant DMR-1123894.  A.C. gratefully acknowledges funding from the Alexander von Humboldt Foundation within the
Feodor-Lynen Fellowship program.  T.C.B. was supported in part by the U.S.~Department of Energy, Office of Science under Contract
No.~DE-AC05-06OR23100.  This work was carried out in part at the Center for Functional Nanomaterials, Brookhaven National
Laboratory, which is supported by the U.S.~Department of Energy, Office of Basic Energy Sciences under Contract
No.~DE-AC02-98CH10886 (M.S.H).

\textit{Note added}. --- After submission of our paper, several additional experimental and theoretical 
reports on the excitonic properties of monolayer TMDs have appeared~\cite{Zhu-arx,*Ye-arx,*Wang-arx,*Ugeda-arx,*Stroucken-arx,*He2014}.


%

\clearpage
\onecolumngrid

\section{Supplemental Material} 

\appendix

\setcounter{figure}{0}
\makeatletter 
\renewcommand{\thefigure}{S\@arabic\c@figure}
\makeatother
\renewcommand{\theequation}{S\arabic{equation}}

\section{Experimental methods}

Spatial images of the samples were obtained using an optical microscope (Nikon Eclipse ME600) with a 100x objective and a CCD camera for detection.
Photoluminescence~(PL) and Raman measurements were performed in in a commercial Raman microscope (Renishaw inVia) under ambient conditions at room
temperature with the respective spectral resolutions of 1 meV and 1 cm$^{-1}$.  The spectra were calibrated using the Si Raman line at 520 cm$^{-1}$.
A continuous-wave solid-state laser with a central wavelength of 532 nm was used for excitation.  The power was set to 50 $\mu$W and the laser was
focused to a 1 $\mu$m diameter spot.

For the reflectance contrast measurements, broadband radiation from a tungsten quartz halogen source was focused on the sample by a 40$\times$
objective, yielding a spot of about 2 $\mu$m in diameter.  The reflected signal was collected and analyzed with a grating spectrometer and a liquid
nitrogen cooled CCD.  The measurements were performed both at room temperature and at 20~K.  The data were oversampled with 10 pixels corresponding to
the experimental resolution of the setup of about 10~meV.  Normalizing the reflectance from the sample to that of the bare substrate, we obtain the
reflection contrast with a signal-to-noise ratio (SNR) of about 10$^3$.  We took advantage of the oversampling to average the derivative spectra over
an interval of $\Delta=10$ pixels.  This further improved the SNR without any significant impact on the spectral resolution.  We explicitly confirmed
that an increase or decrease of $\Delta$ by a factor of two did not affect the measured peak positions.  The latter were obtained by taking the
respective points of inflection, corresponding to the zero-crossings of the second derivative.  We also carried out simulations of the full optical
response of the sample and substrate, including the influence of multiple reflections from the substrate and oxide layer, as well as the WS$_2$
response. The exciton transition energies inferred from this procedure agreed within experimental error with those obtained from the simple point of
inflection analysis (see section ``Simulation of the low-temperature reflection contrast'' for details). 

In general, not only the high SNR of the experiment but also the quality of the samples is crucial for the observation of higher excited states reported in the manuscript.
Only the samples with sufficiently narrow line-widths of the excitonic transitions allow for the detection of weaker resonances above the ground state, which can be further facilitated by performing the measurements at low temperatures.
The line-width of the higher transitions is found to scale with the line-width of the ground state, yet it is generally significantly broader than the main resonance due to the presence of additional relaxation processes.
Moreover, the residual doping of the samples significantly broadens higher lying exciton states, precluding their observation for samples with high intrinsic doping.
Nevertheless, higher exciton states were observed in four different WS$_2$ monolayers, including room-temperature measurements.

\section{Optical microscopy images}

\begin{figure}[t]
\centering
\includegraphics[width=7cm]{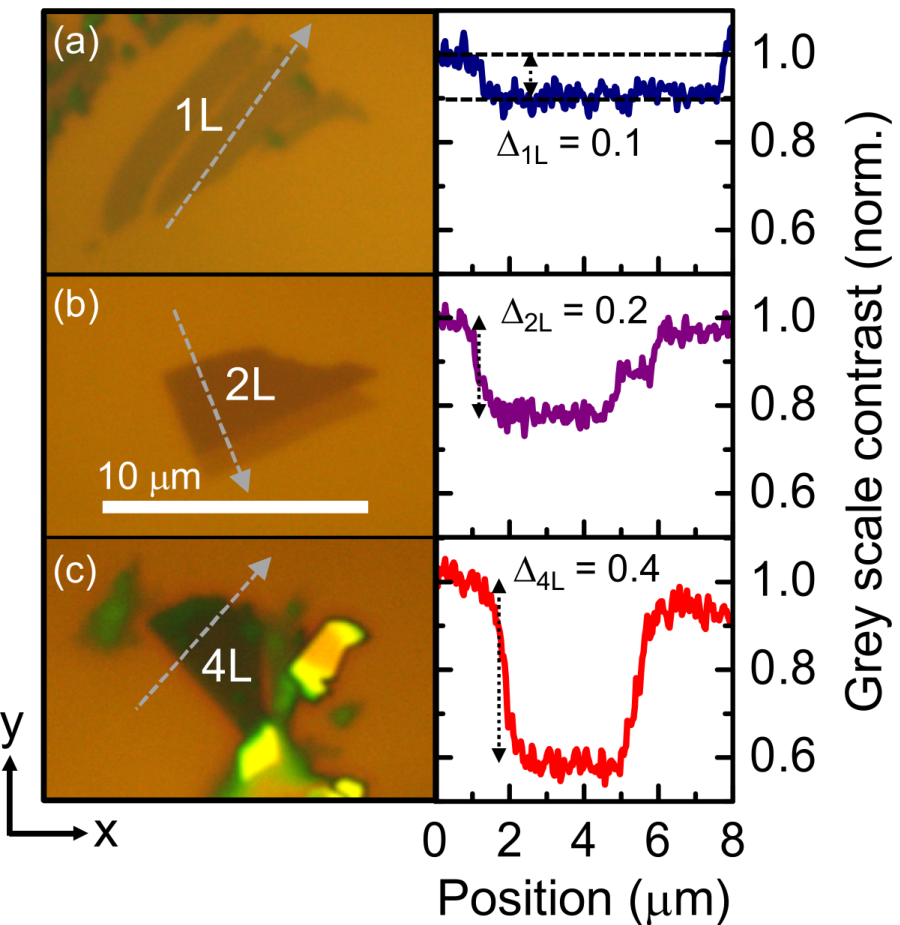}
\caption{
Optical micrographs of the studied WS$_2$ monolayer~(1L), bilayer~(2L), and tetralayer~(4L) samples. 
Exemplary grey-scale contrast profiles are taken along the dashed lines and shown on the right-hand side of the respective images.
The data is normalized to the contrast of the SiO$_2$/Si substrate.
}
\label{fig:micrographs} 
\end{figure}

Figures~\ref{fig:micrographs}(a), (b), and (c) present optical microscopy images of the WS$_2$ monolayer~(1L), bilayer~(2L), and tetralayer~(4L) samples,
respectively.  The average area of the studied flakes is in the range of 20 $\mu$m$^2$, suitable for optical studies using spots with diameters up to
several $\mu$m.  Representative grey scale contrast profiles are taken along the dashed lines and plotted on the right-hand side of the images,
normalized to the contrast of the SiO$_2$/Si substrate.  The discrete steps of about 0.1 observed in the grey-scale contrast correspond to the number
of WS$_2$ layers.

\section{Room-temperature photoluminescence and reflection contrast measurements}
\begin{figure}[h]
\centering
\includegraphics[width=11cm]{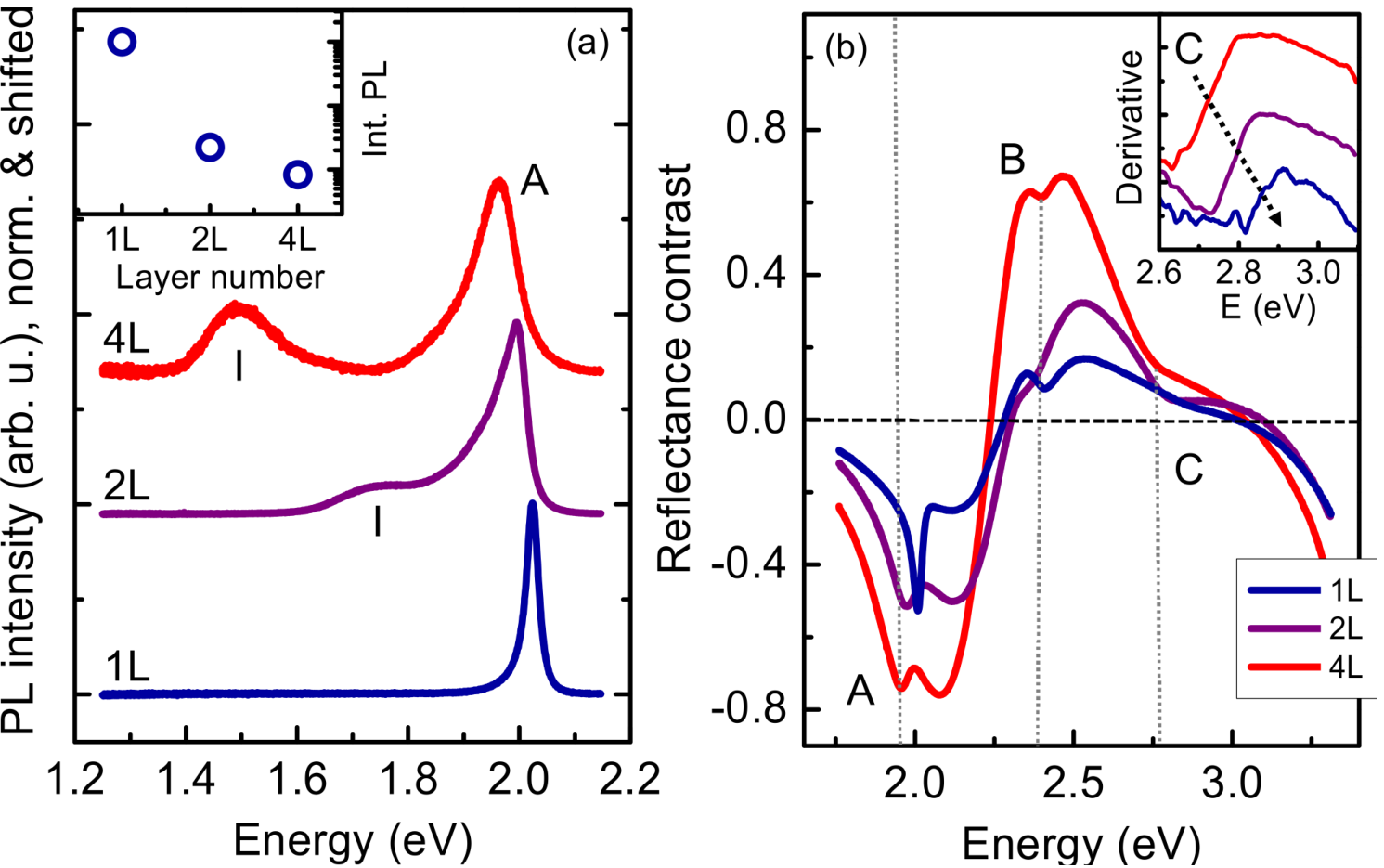}
\caption{(a)~Room-temperature photoluminescence spectra of the WS$_2$ mono- and few-layer samples.  The data are normalized and offset for clarity.
The main $K$-point transition and the indirect band gap are indicated by A and I, respectively.  The overall luminescence yield is plotted in the
inset as function of the layer thickness.  (b)~Corresponding spectra of the reflectance contrast with the spectral positions of the direct transitions
marked by A, B, and C.  The inset shows the smoothed derivatives of the reflectance contrast, normalized and offset for clarity, in the range of the C
resonance.
} 
\label{fig:PL-refl}
\end{figure}
Relying on previous studies of optical properties of the TMDs and the WS$_2$ system in particular~\cite{sup-Mak2010,sup-Cao2012,sup-Tonndorf2013,sup-Zhao2013}, we are
able to identify unambiguously the layer thickness of our samples using a combination of PL and reflection contrast data.  PL spectra of the studied
WS$_2$ samples are shown in Fig.~\ref{fig:PL-refl}(a), normalized and offset for clarity.  The corresponding total luminescence yield is plotted in the inset.
The main $K$-point resonance and the indirect band gap emission are identified by the labels A, and I, respectively.  The emission from the monolayer
is characterized by an increase of the PL intensity by about two orders of magnitude compared to the bilayer and by the absence of an indirect
transition in the luminescence spectra.  The measured positions of the indirect gap at 1.73 eV and 1.49 eV for the bi-and tetralayer, respectively,
agree well with the published energies of 1.72 eV and 1.48 eV from Ref.~\onlinecite{sup-Zhao2013}.  For comparison, the energies for the tri- and
pentalayer were previously found to be 1.53 eV and 1.42 eV~\cite{sup-Zhao2013}.  Also, the red shift of the indirect transition is accompanied by a
further decrease of the emission intensity.

Figure~\ref{fig:PL-refl}(b) shows the measured reflectance contrast spectra for the three samples.  The overall sinusoidal shape of the spectra with values
above and below zero is due to the change in the interference pattern typical for layers supported by SiO$_2$/Si substrates (see
e.g.~Refs.~\onlinecite{sup-Mak2012,sup-Ross2013}).  The spectral positions of the direct transitions are marked by A, B, and C according to the usual labeling
convention~\cite{sup-Wilson1969,sup-Zhao2013}.  The inset shows smoothed derivatives of the reflectance contrast in the range of the C resonance, normalized
and offset for clarity.  The energy positions of the C peak transition of 2.86 eV, 2.78 eV, and 2.72 eV for the mono-, bi-, and tetralayer,
respectively, agree with the previously reported values of 2.83 eV, 2.75 eV, and 2.70 eV~\cite{sup-Zhao2013} within the experimental uncertainty.  The
blue-shift of the A and B resonances, as well as the overall decrease of the reflection contrast with decreasing thickness, further corroborate the
assignment of the layer number.

\section{Raman spectroscopy}

\begin{figure}[h]
\centering 
\includegraphics[width=5.5 cm]{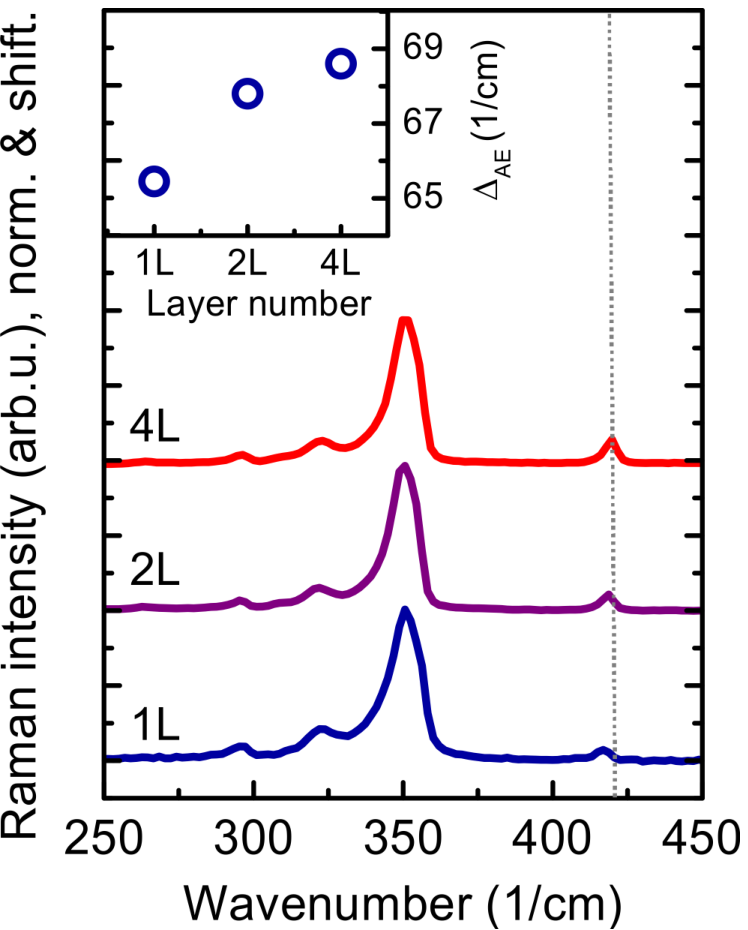}
\caption{Room temperature Raman spectra of the WS$_2$ mono- and few-layer samples.  The spectra are normalized and offset for clarity.  The energy
difference between the A$_{1g}$ and E$^1_{2g}$ modes is plotted in the inset as function of the layer thickness.
} 
\label{fig:raman}
\end{figure}

The Raman spectra for the three samples are plotted in Fig.~\ref{fig:raman}, normalized and offset for clarity.  The observed peaks are identified as
follows: the A$_{1g}$($\Gamma$) optical mode at 417 cm$^{-1}$, E$^1_{2g}$($\Gamma$) optical mode at 354.4 cm$^{-1}$ merged with the second order
acoustical mode 2xLA(M) at 350.6 cm$^{-1}$, and a coupled acoustical-optical mode 2xLA(M)-3xE$^2_{2g}$(M) at 295.4
cm$^{-1}$~\cite{sup-Gutierrez2012,sup-Zhao2013a}.  The precise origin of the resonances at 322 cm$^{-1}$ and 312.5 cm$^{-1}$ is still under
discussion~\cite{sup-Zhao2013a}.  As the layer number increases, the A$_{1g}$($\Gamma$) mode blue shifts by a few wavenumbers and the E$^1_{2g}$($\Gamma$)
and 2xLA(M) modes redshift very gradually, as previously reported~\cite{sup-Gutierrez2012}.  The energy difference between the A$_{1g}$ and E$^1_{2g}$
modes, shown in the inset, thus increases with the layer number, as is typically observed for the TMDs~\cite{sup-Lee2010,sup-Gutierrez2012}.

\section{Simulation of the low-temperature reflection contrast}
\label{simulation}

\begin{figure}[t]
\centering 
\includegraphics[width=15 cm]{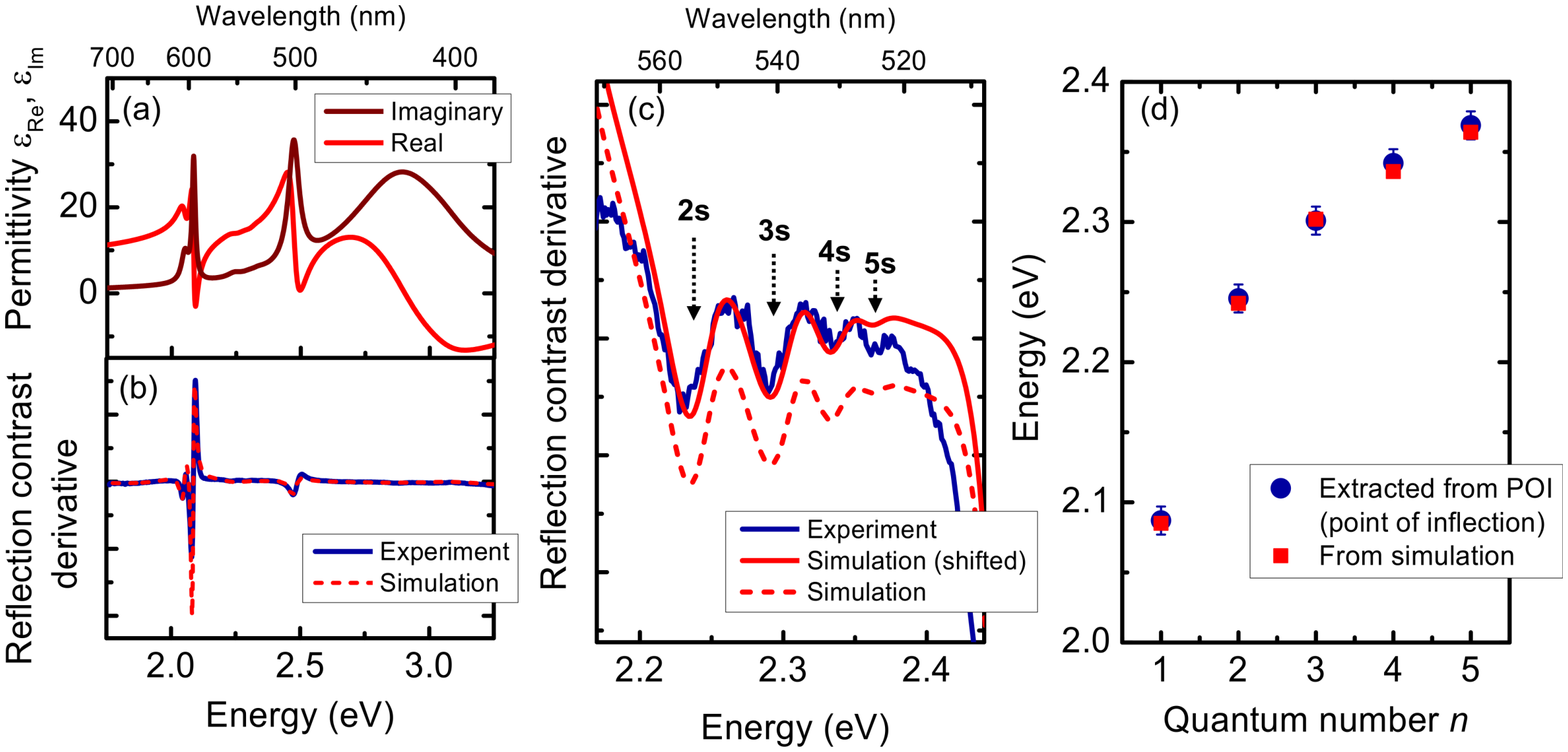}
\caption{Simulations of the optical response based on the transfer-matrix method to account for multiple reflections. 
(a)~The real and imaginary parts of the dielectric function used in the simulation to fit the measured reflectivity of the WS$_2$ monolayer on the
SiO$_2$/Si substrate at temperature of 20 K.  The permittivity at the excitonic resonances is parametrized with multiple Lorentzians.  (b)~Measured
and simulated derivatives of the reflection contrast.  (c)~Magnified derivatives of the reflection contrast in the spectral region of the A-exciton
fine-structure. The simulated data have been also shifted by a constant value for better comparison with the measured lineshape.  (d)~Exciton peak
energies extracted by taking the points of inflection are compared with the values from the simulation.
} 
\label{fig:simulation}
\end{figure}

To verify the energy positions of the exciton peaks in the presence of the interference from the underlying SiO$_2$/Si substrate we simulate the
optical response of the monolayer WS$_2$ at 20~K using the standard transfer-matrix method for evaluating light propagation in planar thin films (see
e.g., Ref.~\onlinecite{sup-Hecht2001}).  The calculations are performed using an open-source software~\cite{sup-Byrnes2012}.  The transfer-matrix technique
simulates the linear optical response of a given multi-layered thin-film system taking into account multiple internal reflections and interference for
layers of arbitrary thicknesses and complex refractive indices.  The wavelength-dependent refractive indices of the SiO$_2$ and Si are taken from
Refs.~\onlinecite{sup-Malitson1965} and \onlinecite{sup-Bass2009}, respectively.  The A, B, and C exciton resonances of the WS$_2$ monolayer as well as the
excited states of the A-exciton (2s, 3s, 4s, and 5s) are parametrized with Lorentzian peak functions; a constant background permittivity is included
to take into account higher-lying transitions.  The peak energies, widths, and spectral weights of the resonances are adjusted to obtain an optimal
overlap of the simulated and measured reflectivity.  Real and imaginary parts of the dielectric function produced by the simulation to provide the
best fit of the experimental data are shown in Fig.~\ref{fig:simulation}(a).  The corresponding derivative of the simulated reflection contrast is plotted in
Fig.~\ref{fig:simulation}(b), together with the measured spectrum.  Fig.~\ref{fig:simulation}(c) presents the same data magnified in the spectral region of the A-exciton
Rydberg series.  The simulated data have been additionally shifted vertically by a constant value for better comparison of the lineshapes.  The
A-exciton peak energies produced by the simulation are plotted in Fig.~\ref{fig:simulation}(d) and compared with the peak positions obtained by taking the points
of inflection~(POI) in the measured reflection contrast derivative.  The approaches are in agreement within the experimental error, as indicated by
the error bars for the POI values.  These findings corroborate the initial assumption that the interference from the multi-layered substrate does not
significantly affect the measured energies of the spectrally narrow exciton resonances.   

\section{Theoretical modeling and exciton properties}

As discussed in the text, we use a microscopic exciton Hamiltonian ($\mathbf{r} = (x,y)$ is the 2D in-plane coordinate)
\begin{equation}
H = -\frac{\hbar^2 \nabla_{\mathbf{r}}^2}{2\mu} + V_{eh}(r)
\end{equation}
where $\mu_{xy} = 1/(m_e^{-1}+m_h^{-1})$ is the
exciton reduced mass in the 2D $xy$ plane, which for WS$_2$ has been previously calculated to be in the range
of 0.15--0.22 $m_0$~\cite{sup-Xiao2012,sup-Ramasubramaniam2012,sup-Shi2013}, depending on the level of theory used ($m_0$ is the rest mass of
the electron). Here we use $\mu = 0.16\ m_0$, but similar results could be obtained with any value in this range.
The screened electron-hole interaction is given in reciprocal-space by
\begin{equation}
V_{eh}(q) = -\frac{2\pi e^2}{q\varepsilon(q)} = -\frac{2\pi e^2}{q(1+r_0 q)}
\end{equation}
or in real-space by
\begin{equation}
V_{eh}(r) = -\frac{\pi e^2}{2r_0} \left[ H_0\left(\frac{r}{r_0}\right) - Y_0\left(\frac{r}{r_0}\right) \right] 
\end{equation} 
where $H_0$ and $Y_0$ are Struve and Bessel functions.  The eigenvalues and eigenfunctions of the Hamiltonian are
calculated numerically by diagonalization on a one-dimensional real-space grid.

The above interaction has been derived by Keldysh
for quasi-2D semiconductors~\cite{sup-Keldysh1979} and by Cudazzo et al.~for strictly 2D semiconductors~\cite{sup-Cudazzo2011}.
In these derivations, the screening lengthscale $r_0$ is given by $r_0 = d \varepsilon / 2$ and $r_0 = 2 \pi \chi_{2D}$,
respectively, where $d$ is the layer thickness, $\varepsilon$ is an isotropic macroscopic dielectric constant, and $\chi_{2D}$
is the 2D polarizability. We have recently applied such a formalism to the family of TMDs to investigate the binding energy
of both neutral excitons and charged trions~\cite{sup-Berkelbach2013}, and found an approximate equivalence between these two definitions.
Such an interaction can be understood as the large wavelength (small $q$) approximation to the electrostatic potential
of a charge inside a layer of thickness $d$ and dielectric constant $\varepsilon$.  Retaining the full $q$-dependence of the dielectric
 function $\varepsilon(q)$ via an image-charge
solution -- as has been done in other works on quantum wells~\cite{sup-Hanamura1988}, inorganic-organic perovskites~\cite{sup-Tanaka2005},
and TMDs~\cite{sup-Zhang2014} -- was not found to modify the results presented here.
Using \textit{ab initio} calculations carried out based on density functional theory and the random phase approximation (DFT+RPA), as
described in detail in Ref.~\onlinecite{sup-Berkelbach2013}, we would predict a screening length of $r_0 = 38$ \AA\ for intrinsic,
suspended WS$_2$.  This value of $r_0$ is roughly half as large as that used in the present paper, and accordingly yields a larger
exciton binding energy of 0.50 eV, which is relevant for comparisons with fully \textit{ab initio} calculations or with future measurements on suspended samples.  
A variational calculation gives a predicted lower bound on the WS$_2$ trion binding energy of
26 meV (15 meV) for the smaller (larger) value of $r_0$, to be compared with our own preliminary experimental estimate of about 30 meV.

\begin{figure}[t]
\centering 
\includegraphics[width=9 cm]{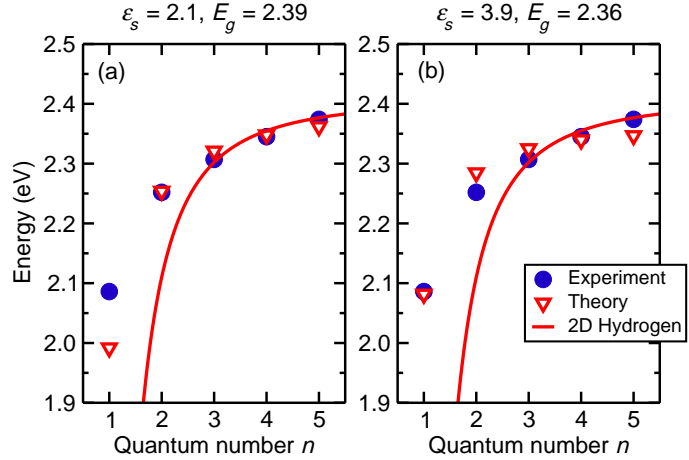}
\caption{Predicted transition energies of monolayer WS$_2$, explicitly accounting for screening due to the SiO$_2$ substrate via Eq.~\ref{eq:int-subst}.
We have considered use of the relative permittivity evaluated at the optical frequency, $\varepsilon_s = 2.1$ (a) and evaluated at zero frequency,
$\varepsilon_s = 3.9$ (b). In each case the band-gap of the theoretical result has been adjusted to give the best agreement with experiment.
The 2D hydrogen model result is unchanged from the result in the main text.
} 
\label{fig:theory-subst}
\end{figure}

We attribute the discrepancy in the screening length $r_0$ to a combination of substrate and local-field effects, the latter of which
cannot be captured by the long wavelength theory outlined above.  Specifically, we envisage an effective higher material
polarizability at short- to mid-range length scales.
Preliminary calculations based on a more detailed classical electrostatic model and based on \textit{ab initio} screening calculations 
of the full dielectric matrix support this hypothesis.  Substrate screening effects can be approximately included in our theory with the 
modified interaction~\cite{sup-Keldysh1979}
\begin{equation}\label{eq:int-subst}
V_{eh}(r) = -\frac{\pi e^2}{2r_0} \left[ H_0\left(\frac{(1+\varepsilon_s)r}{2r_0}\right) - Y_0\left(\frac{(1+\varepsilon_s)r}{2r_0}\right) \right], 
\end{equation}
where $r_0$ is the screening length in the absence of a substrate and $\varepsilon_s$ is the dielectric constant of the substrate.  This substrate screening
has subtle, nontrivial effects on the effective electron-hole interaction: in addition to reducing the overall strength of the interaction,
the renormalization of the screening length reduces the range over which the potential exhibits a logarithmic form.  
We have calculated the binding energy using the above interaction, with two different possible dielectric constants for the SiO$_2$ substrate:
the relative permittivity at the optical frequency, $\varepsilon_s = 2.1$, and at zero frequency, $\varepsilon_s = 3.9$. The results are shown in 
Fig.~\ref{fig:theory-subst}, panels (a) and (b) respectively.  The screening length $r_0$ has been fixed to the \textit{ab initio} calculated value of
$r_0 =$ 38 \AA, such that \textit{the band gap $E_g$ is the only free parameter}. For these two cases, we find 1s exciton binding energies
of 0.40 and 0.28 eV, which are in reasonable agreement with the value of 0.32 eV obtained in the main text.
Both models underestimate the energy of the measured $n = 5$ transition, however this experimental data point carries the largest degree of uncertainty.
Notably, each theoretically-predicted excitonic Rydberg series shown in Fig.~\ref{fig:theory-subst} is still significantly non-hydrogenic.
Together these findings provide a strong indication that substrate effects can be partially accounted for by a phenomenologically 
increased screening length as is done in the main text.

Compared to other theoretical treatments of excitons in TMDs, we point out that two other model-based studies by 
Zhang et al.~\cite{sup-Zhang2014} and by Bergh{\"a}user and Malic~\cite{sup-Berghauser2014} have employed
our treatment of 2D dielectric behavior~\cite{sup-Berkelbach2013}, which follows Keldysh~\cite{sup-Keldysh1979} and 
Cudazzo et al.~\cite{sup-Cudazzo2011}. These two recent works are in good qualitative agreement with our own: the former focused on the
ground-state exciton and found a binding energy of 0.28--0.33 eV, while the latter identified the existence of an s-type excitonic
Rydberg series, but did not comment on its non-hydrogenic behavior.  Fully \textit{ab initio} studies based on the
GW+BSE approach~\cite{sup-Ramasubramaniam2012,sup-Komsa2012,sup-Shi2013,sup-Qiu2013} do not require a 
model-based form for the dielectric function; they have found similar characteristics for the spectrum of excited states, although
with higher overall binding energies.

For the binding energy of \textit{bulk} WS$_2$, we employ the anisotropic exciton Hamiltonian
\begin{equation}
H_{\mathrm{bulk}} = \frac{p_x^2+p_y^2}{2\mu_{xy}} + \frac{p_z^2}{2\mu_{z}} - \frac{e^2}{\left[ \varepsilon_{z} \varepsilon_{xy} (x^2+y^2) + \varepsilon_{xy}^2 z^2 \right]^{1/2}}
\end{equation}
which can be transformed via the change of variable $z \rightarrow (\mu_{z}/\mu_{xy})^{1/2}z$ into
\begin{equation}
\widetilde{H}_{\mathrm{bulk}} = \frac{p_x^2+p_y^2+p_z^2}{2\mu_{xy}} - \frac{e^2}{\left[ \varepsilon_{z} \varepsilon_{xy} (x^2+y^2+\gamma z^2) \right]^{1/2}}
\end{equation}
where the anisotropy parameter is defined as $\gamma = \varepsilon_{xy} \mu_{xy} / (\varepsilon_{z} \mu_{z})$.
Using the values calculated by DFT+RPA ($\mu_{xy} = 0.16\ m_0,\ \mu_z = 1.2\ m_0,\ \varepsilon_{xy} = 13,\ \varepsilon_z = 6.3$),
we find $\gamma \approx 0.3$. Although not analytically tractable, the eigenvalues of this anisotropic Hamiltonian have been
numerically calculated previously~\cite{sup-Baldereschi1970}.  Using the results of Ref.~\onlinecite{sup-Baldereschi1970}
with $\gamma = 0.3$, we find a binding energy of 45 meV for bulk WS$_2$.

\section{Computational details}

All single-particle electronic structure calculations (for the extraction of effective masses and RPA screening parameters) were
carried out for the experimental crystal structure of WS$_2$ using the \textsc{Quantum Espresso}~\cite{sup-Giannozzi2009} and
\textsc{BerkeleyGW}~\cite{sup-Deslippe2012} software packages.  DFT calculations were performed with $12\times12\times3$ and $12\times12\times1$ $k$-point
grids for the bulk and monolayer, respectively, using the PBE exchange-correlation functional~\cite{sup-Perdew1996}, norm-conserving pseudopotentials, and
a plane-wave cutoff of 40 Ry ($\sim 550$ eV).  Subsequent RPA calculations were utilized to extract the dielectric properties, summing over
approximately 50 unoccupied bands per layer.

\end{document}